\documentclass[twocolumn,aps,showpacs,amsmath]{revtex4-1}
\usepackage[latin9]{inputenc}
\usepackage{array}
\usepackage{amsmath}
\usepackage{graphicx}
\usepackage{esint}

\makeatletter

\providecommand{\tabularnewline}{\\}

\usepackage{color}

\newcommand{\be}{\begin{equation}}
\newcommand{\ee}{  \end{equation}}
\newcommand{\ba}{\begin{eqnarray}}
\newcommand{\ea}{  \end{eqnarray}}
\newcommand{\ket}[1]{\left|#1\right>}

\makeatother

\begin{document}
\title{Optimal photon energies for initialization of hybrid spin quantum
registers of NV centers in diamond}
\author{K. Rama Koteswara Rao$^{1,2}$, Yihua Wang$^1$, Jingfu Zhang$^1$, and Dieter Suter$^1$}
\affiliation{$^1$Fakult{ä}t Physik, Technische Universit{ä}t Dortmund, D-44221
Dortmund, Germany\\
$^2$Department of Physics, Bennett University, Greater Noida 201310, India}
\date{\today}
\begin{abstract}
Initializing quantum registers with high fidelity is a fundamental
precondition for many applications like quantum information processing
and sensing. The electronic and nuclear spins of a Nitrogen-Vacancy
(NV) center in diamond form an interesting hybrid quantum register
that can be initialized by a combination of laser, microwave, and
radio-frequency pulses. However, the laser illumination, which is
necessary for achieving electron spin polarization, also has the unwanted
side-effect of depolarizing the nuclear spin. Here, we study how the
depolarization dynamics of the $^{14}$N nuclear spin depends on the
laser wavelength. We show experimentally that excitation with an orange
laser (594 nm) causes significantly less nuclear spin depolarization
compared to the green laser (532 nm) typically used for excitation
and hence leads to higher nuclear spin polarization. This could be
because orange light excitation inhibits ionization of NV$^{0}$ into
NV$^{-}$ and therefore suppresses one source of noise acting on the
nuclear spin.
\end{abstract}
\maketitle

\section{Introduction}

Nitrogen-Vacancy (NV) centers in diamond have interesting properties
for spin based quantum information processing and nano-scale Nuclear
Magnetic Resonance (NMR) spectroscopy and imaging \cite{Chil2006Sci,Lukin2007Sci,JW2013Sci,JW2014Nat,RevDoherty,RevChildress,RevWalsworth,RevDegen}.
Nuclear spins coupled to NV centers are useful resources for these
applications. They can be used as qubits in a hybrid quantum register
\cite{Lukin2007Sci,JW2008Sci,JW2014Nat,Han2014NatNano,ZhangPRL2015}
or as long-lived memories to store quantum states of electron spins
\cite{Awsch2011NatPhy,Lukin2012Sci,Shim2013}. They can also be used
as a channel for transferring polarization between electron spins
of NV centers and nuclear spins of the bulk (remotely coupled) in
hyperpolarization experiments \cite{Frydman2013hyperpol,Alvarez2015,My_hyperpol2018,AjoyDNP2018PNAS}.
Initializing or polarizing nuclear spins is an essential part of these
experiments.

Electron spins of NV centers can be polarized near completely by optical
pumping. However, this process does not automatically lead to the
polarization of nuclear spins coupled to NV centers. Different methods
for polarizing these nuclear spins have been discussed in the literature
\cite{JacquesESLAC,JW2010Sci,Shim2013,PaglieroAPL2014,Tanmoy2017pol,DuChopped2019}.
One of them makes use of a level anti-crossing in the excited state
which occurs in a magnetic field of 51.2 mT oriented along the NV
axis \cite{JacquesESLAC}. Anti-crossing at this magnetic field causes
mixing between the electron and nuclear spin states of the excited
state and leads to polarization of both the electron and nuclear spins
under optical pumping. This method was successfully used to polarize
the $^{14}$N nuclear spin of an NV center and a $^{13}$C nuclear
spin of the first coordination shell. However, this method does not
lead to good polarization of other $^{13}$C nuclear spins \cite{Jacques13Chyperfine},
and it is only applicable at one specific strength and orientation
of the magnetic field. Another interesting method to initialize nuclear
spins of NV centers is through single shot readout \cite{JW2010Sci}.
This method has been used to initialize the $^{14}$N nuclear spin
of an NV center and also a specific $^{13}$C nuclear spin \cite{Jacques2NspinIni}.
However, this method also requires a strong static magnetic field
compared to the transverse components of the hyperfine interaction
\cite{JW2010Sci}.

A more general method to polarize nuclear spins coupled to NV centers
is to apply a sequence of microwave (MW), radio-frequency (RF), and
laser pulses \cite{Shim2013,PaglieroAPL2014,Tanmoy2017pol,DuChopped2019}.
The basic idea of this method is to first polarize the electron spin
and then transfer this polarization to a nuclear spin coupled to it,
using MW and RF pulses. The electron spin, which is then left in a
mixed state, can be repolarized by a second laser pulse. However,
this laser pulse causes depolarization of the nuclear spin and the
degree of depolarization depends on the power and duration of the
laser pulse. One possible source of nuclear spin depolarization is
the ionization of the NV center during the laser pulse.

In this work, we study the depolarization dynamics of the $^{14}$N
nuclear spin of an NV center for different wavelengths of laser illumination
including 532 and 594 nm. The absorption cross-section of the NV center
at both wavelengths is roughly the same, but they cause very different
ionization rates: The 532 nm photons can ionize NV$^{0}$ into NV$^{-}$
and vice-versa by a two-photon process, while the 594 nm photons can
ionize NV$^{-}$ into NV$^{0}$ but the probability rate from NV$^{0}$
into NV$^{-}$ is very small at this wavelength \cite{Nabeel_Ion}.
Here, we show experimentally that the depolarization rate of the $^{14}$N
nuclear spin is significantly lower for 594 nm irradiation than for
532 nm, while the polarizing rate of the electron spin is roughly
the same for both wavelengths, resulting in higher nuclear spin polarization
under 594 nm excitation. Although all the wavelengths between 575
and 637 nm significantly inhibit ionization of NV$^{0}$ into NV$^{-}$,
wavelengths around 594 nm are optimal for the present purpose because
of the high absorption cross-section of NV$^{-}$ charge state at
these wavelengths \cite{Nabeel_Ion}.

This paper is arranged as follows. In Section II, we describe the
method for polarizing the nuclear spin and the differences between
the green and orange light excitation of an NV center. In Section
III, we give the details of our experiment and discuss the results
and in Section IV, we conclude.

\section{Polarization method}

We consider polarizing the single $^{14}$N ($I=1$) nuclear spin
coupled to the electronic spin $S=1$. The Hamiltonian of such a system
interacting with a static magnetic field aligned along the NV axis
can be written as

\begin{align}
{\cal H}= & DS_{z}^{2}+\gamma_{e}BS_{z}+\gamma_{n}BI_{z}+PI_{z}^{2}\nonumber \\
 & +A_{\parallel}S_{z}I_{z}+A_{\perp}(S_{x}I_{x}+S_{y}I_{y}).\label{eq:Hamiltonian}
\end{align}
Here, $S_{\alpha}$ and $I_{\alpha}$ represent the $\alpha$-components
of the spin angular momenta of the electronic and nuclear spins respectively,
and $\gamma_{e}$ and $\gamma_{n}$ are their respective gyro-magnetic
ratios. $D=2870$ MHz and $P=-4.95$ MHz \cite{Bajaj14Nquadrapole}
are the zero-field splitting of the electron spin and the quadrupole
splitting of the $^{14}$N nucleus, measured in frequency units. $B$
represents the strength of the static magnetic field, and $A_{\parallel}=-2.3$
MHz and $A_{\perp}=-2.6$ MHz \cite{Mansion14Nhyperfine,Felton2009,PaolaPRB2015}
are the components of the hyperfine interaction along the NV axis
and perpendicular to it.

\begin{figure}
\includegraphics[width=8.8cm]{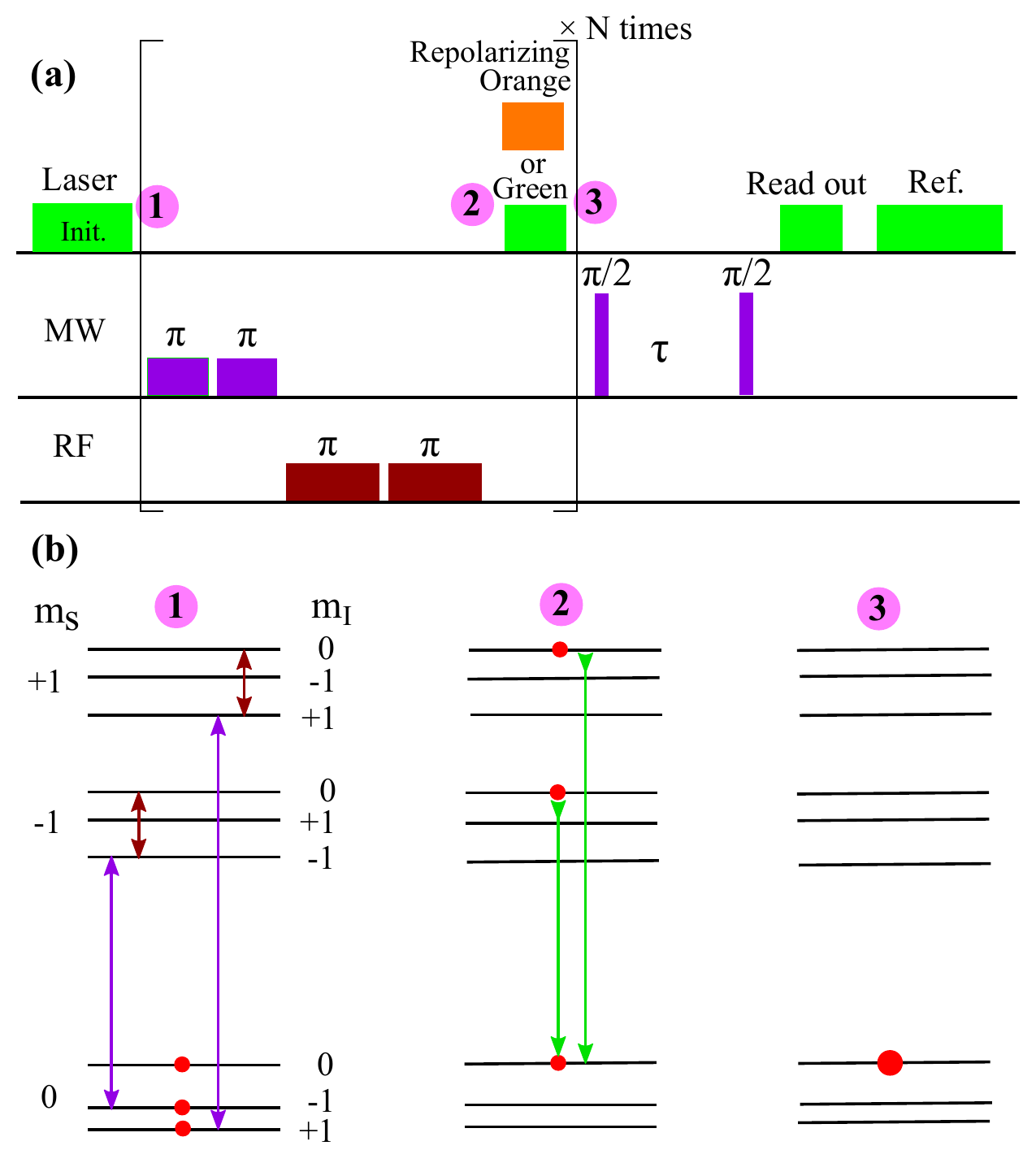} \caption{Schematic representation of the polarization method. (a) Pulse sequence;
green or orange rectangle in the first line represent corresponding
laser pulses. The rectangles in the second and third row represent
MW and RF pulses applied at resonance to the electronic and nuclear
spin transitions respectively. The $\pi/2-\tau-\pi/2$ sequence in
the second line is used to measure the free-induction decay of the
electron spin. (b) Energy level diagram and corresponding populations
at different stages of the pulse sequence}
\label{Pulseseq}
\end{figure}

Fig. \ref{Pulseseq} shows a schematic representation of the method
for polarizing the two spins. Fig. \ref{Pulseseq}(a) shows the pulse
sequence. The first laser pulse polarizes the electron spin into the
$m_{s}=0$ state, but this leaves the nuclear spin in a mixed state
as illustrated in Fig. \ref{Pulseseq}(b). The polarization of the
electron spin can be transferred to the nuclear spin by applying two
electron spin transition selective MW $\pi$ pulses followed by another
two nuclear spin transition selective RF $\pi$ pulses. Now, the nuclear
spin is fully polarized, but the electron spin is in the completely
mixed state. To repolarize the electron spin, we need to apply another
laser pulse. However, this laser pulse causes partial depolarization
of the nuclear spin \cite{Tanmoy2017pol}. The amount of depolarization
depends on the intensity and duration of the laser pulse. There may
be different sources of noise that cause depolarization of the nuclear
spin. One possible source is the ionization of NV$^{-}$ into NV$^{0}$
and vice-versa during optical illumination: the electronic spin of
NV$^{0}$ is $S=1/2$ and its hyperfine interaction is different from
that of NV$^{-}$.

All the experiments that are reported so far use green light (532
or 520 nm) to initialize and repolarize the NV center. It is known
that under green light illumination, the charge state of an NV center
flips between the NV$^{-}$ and NV\textsuperscript{0} states with
an average distribution of the NV$^{-}$ and NV$^{0}$ populations
being 70 and 30 \% respectively \cite{Nabeel_Ion}. This implies that
by the end of the initialization laser pulse, the center would be
in the NV$^{0}$ state with 30\% probability. In this case, the subsequent
MW and RF pulses have no effect on the spin. However, the repolarizing
laser pulse can convert it into NV$^{-}$ and this state contributes
to the observed signal. Since polarization transfer does not occur
for this, signal contribution from it results in reduced polarization
of the nuclear spin.

\begin{figure}
\includegraphics[width=5cm]{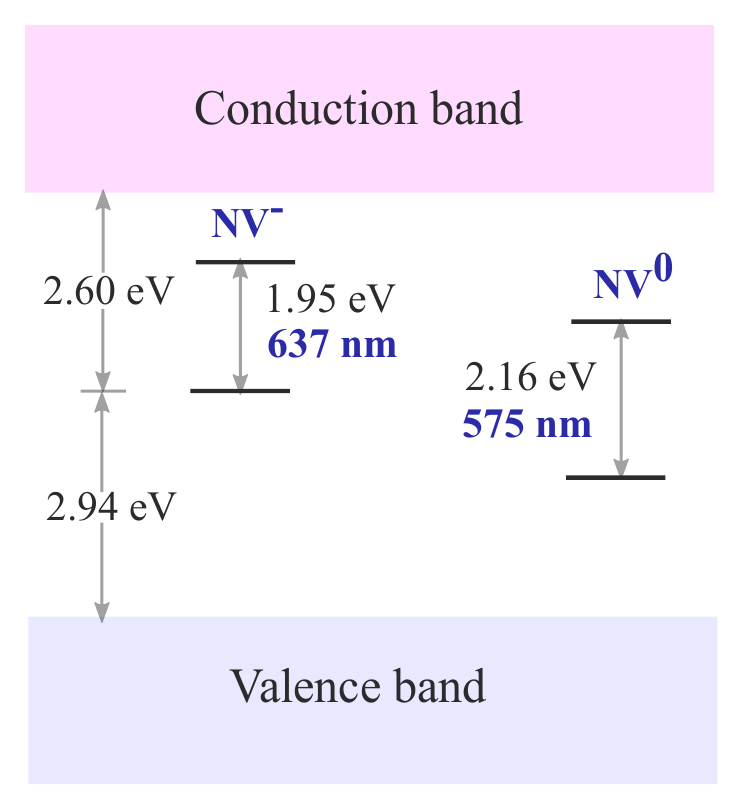} \caption{Schematic representation of the energy levels of an NV center for
its negative and neutral charge states in the band gap of diamond.}

\label{NVenglvls}
\end{figure}

In order to eliminate this depolarization channel, we therefore change
the protocol: for the repolarization laser pulse, we use an orange
laser, operating at 594 nm, instead of the conventional green laser.
As we show in the following, this leads to a significant reduction
of the depolarization process and results in higher nuclear spin polarization.
The absorption cross-section of NV$^{-}$ is roughly the same for
both lasers, but the orange light does not result in ionization of
NV$^{0}$ into NV$^{-}$ \cite{Nabeel_Ion}. Accordingly, it avoids
signal contribution if the center's charge state is changed during
the repolarizing pulse. An important point to note here is that a
single NV center, under green light readout, generates very little
fluorescence attributable to its NV$^{0}$ state \cite{Nabeel_Ion}.

Fig. \ref{NVenglvls} illustrates the relative positions of the energy
levels of NV$^{-}$ and NV$^{0}$ in the band gap of diamond. The
Zero-Phonon Lines (ZPL) of the NV$^{-}$ and NV$^{0}$ charge states
occur at 637 and 575 nm, respectively. This implies that photons of
wavelength 532 or 520 nm can excite both charge states and also can
ionize one into the other. However, photons of wavelength 594 nm can
excite the NV$^{-}$ state, but not NV$^{0}$. Since photo-induced
ionization of an NV center at the mentioned wavelengths by a two-photon
process necessitates its excitation from the ground to the excited
state, 594 nm light can only ionize NV$^{-}$ into NV$^{0}$, but
not the other way \cite{Nabeel_Ion}. This is true for all wavelengths
between 575 and 637 nm. Since the NV$^{-}$ charge state has a high
absorption cross-section around 590 nm, allowing fast polarization
of its electron spin \cite{Nabeel_Ion}, excitation with a wavelength
around 590 nm should be optimal for the present purpose.

\section{Experimental Results}

All experiments have been performed on a single NV center from a 99.99
\% $^{12}$C enriched bulk diamond sample with a nitrogen concentration
of $<$ 5 ppb. These experiments were also repeated on another center
from the same sample and the results are very similar. The setup used
for these experiments was based on a home-built optical confocal microscope
equipped with 520, 532 and 594 nm lasers for optical excitation of
the NV center and MW and RF electronics for resonant excitation of
electron and nuclear spins. The fluorescence of the NV center was
collected through a 605 nm dichroic mirror followed by a 594 nm long
pass filter. An electromagnet was used to apply a static magnetic
field of 2.8 mT oriented along the NV axis.

The pulse sequence given in Fig. \ref{Pulseseq} was implemented in
the following way. A 4 $\mu$s long 520 or 532 nm laser pulse was
applied to initialize the charge and spin states of the NV center
into the NV$^{-}$, $m_{s}=0$ states. The following MW $\pi$ pulses
were applied to the transitions $\ket{m_{s},m_{I}}=\ket{0,-1}$ $\longleftrightarrow$
$\ket{-1,-1}$, and $\ket{0,+1}$ $\longleftrightarrow$ $\ket{+1,+1}$,
whose frequencies were 2789.13 and 2947.42 MHz respectively. The duration
of each of these pulses was 1 $\mu$s. The RF $\pi$ pulses were applied
to the nuclear spin transitions, $\ket{-1,-1}$ $\longleftrightarrow$
$\ket{-1,0}$, $\ket{+1,+1}$ $\longleftrightarrow$ $\ket{+1,0}$,
whose frequencies were 7.1064 and 7.1226 MHz respectively, and the
duration of each of these pulses was 62 $\mu$s. The repolarizing
laser pulse was derived from the 520, 532 or 594 nm laser. Then, an
electron spin free-induction decay (FID) was measured by applying
the Ramsey sequence ($\pi/2-\tau-\pi/2$) between the $m_{s}=0$ and
$-1$ subspaces followed by a 400 ns readout laser pulse. Here, the
$\pi/2$ MW pulses were non-selective and excited all allowed transitions
between these subspaces. Since the repolarizing laser pulse brings
the populations of the $m_{s}=-1$ and $+1$ subspaces into the $m_{s}=0$
subspace, the intensities of the spectral lines obtained by Fourier
transforming the free-induction decay represent populations of the
corresponding nuclear spin sub-levels. For the experiments involving
532 and 594 nm repolarizing laser pulses, the initialization and readout
pulses were derived from the 532 nm laser and for those involving
520 nm repolarizing laser pulse, the same 520 nm laser was used for
initialization and readout.

\begin{figure}
\includegraphics[width=9cm]{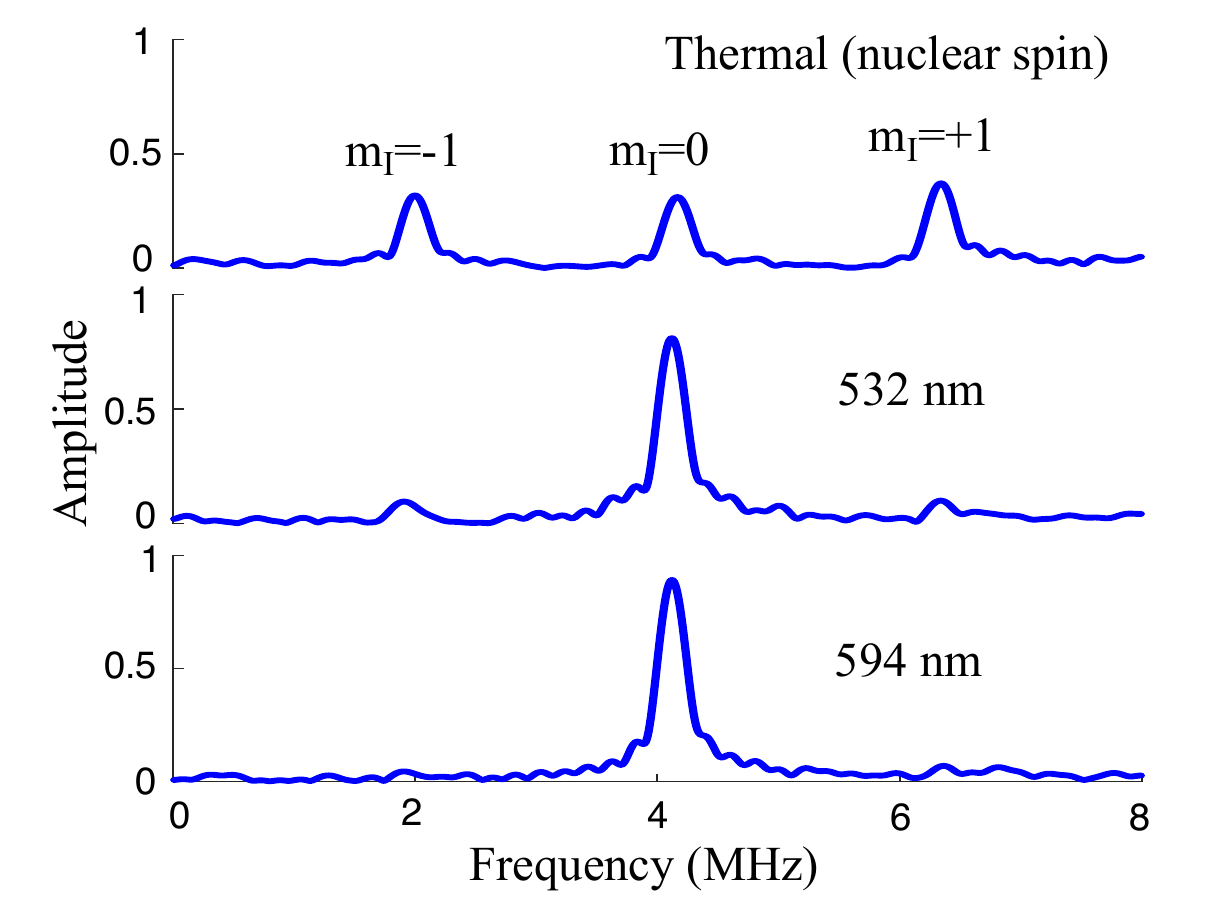} \caption{Fourier transforms of electron spin FIDs measured between the $m_{s}=0$
and $-1$ subspaces. The top row corresponds to the spectrum with
thermal nuclear spin polarization, the middle and bottom rows correspond
to the spectra obtained after the nuclear spin polarizing pulse sequence
of Fig. \ref{Pulseseq} with $N=4$ (cycles) for 532 nm (duration,
500 ns) and 594 nm (700 ns) illumination respectively.}
\label{Spectra}
\end{figure}

The spectra obtained by applying the pulse sequence of Fig. \ref{Pulseseq}
with $N=4$ cycles of polarization transfer and repolarizing pulses
for 532 and 594 nm repolarizing illumination are shown in Fig. \ref{Spectra},
together with a spectrum showing thermal nuclear spin polarization.
This spectrum was obtained by applying the initializing laser pulse
followed directly by the Ramsey sequence (\textit{i.e.} $N=0$). It
contains three lines corresponding to the three $^{14}$N nuclear
spin states, $m_{I}=-1$, $0$, and $+1$, which are split by the
hyperfine coupling. These three lines have roughly equal amplitude
which implies that the nuclear spin is in the maximally mixed state
after the initializing pulse. The spectrum corresponding to the 532
nm repolarizing illumination shows significantly decreased outer peaks
and an increased central peak, which implies that the population of
$m_{I}=-1$ and $m_{I}=+1$ states is transferred to the $m_{I}=0$
state. The spectrum corresponding to the 594 nm repolarizing illumination
shows almost no outer peaks and a strong central peak. We calculate
the nuclear spin polarization ($p$) by writing its density matrix
as $p\left|0\right\rangle \left\langle 0\right|+(1-p)I$, where $I$
is the $3\times3$ identity matrix. From the spectra, we obtain $p$
as $76.3\ (\pm1.9)\ \%$ and $89.0\ (\pm2.7)\ \%$ for the 532 and
594 nm repolarizing illumination respectively.

\begin{figure}
\includegraphics[width=9cm]{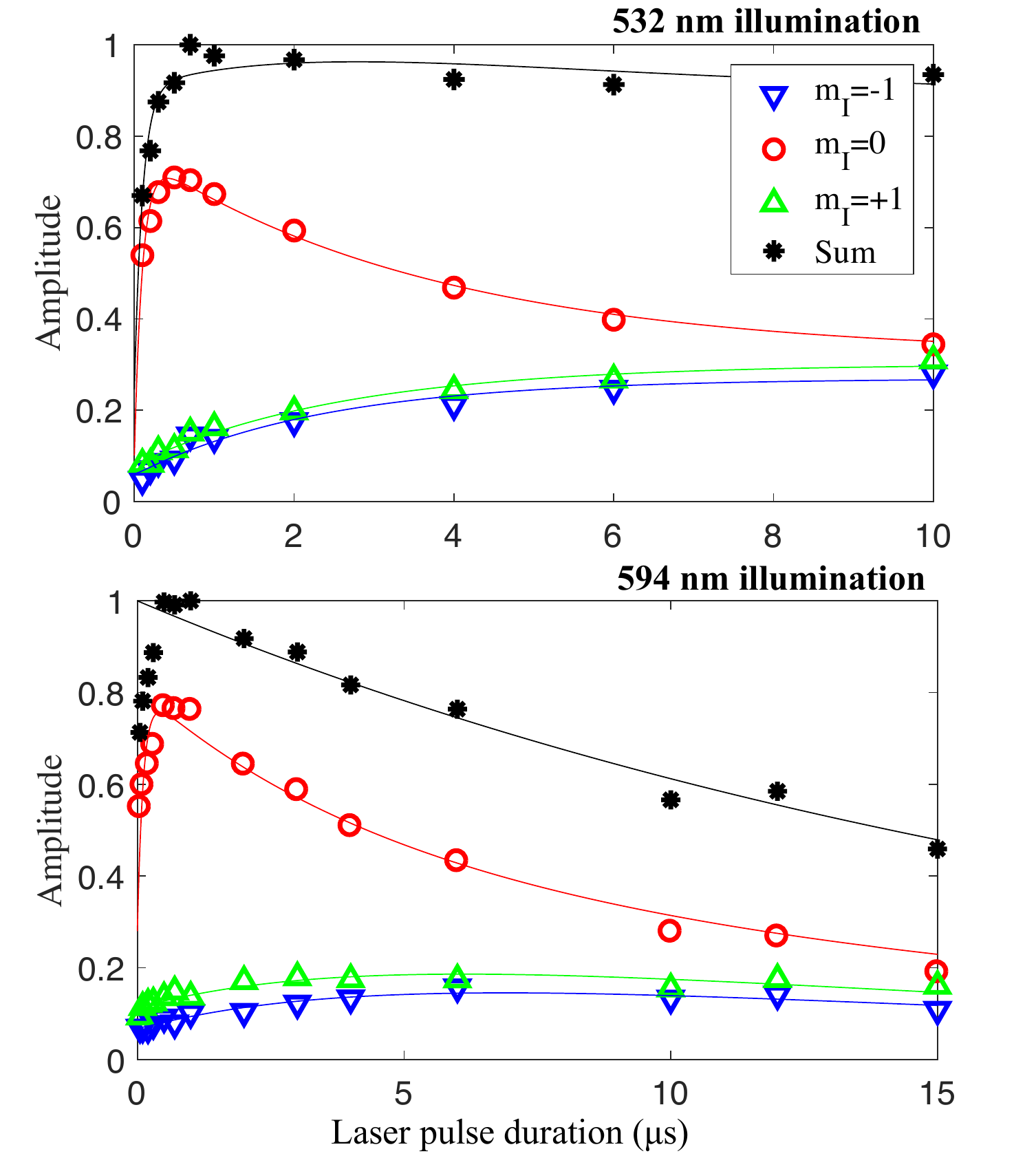} \caption{Dynamics of electron spin polarization and nuclear spin depolarization
as a function of the duration of the repolarizing laser pulse. Circles,
lower, and upper triangles represent experimental data corresponding
to the populations of the $m_{I}=0$, $-1$, and $+1$ states of the
$^{14}$N nuclear spin respectively, and asterisks represent the total
population. The experimental data are compared to the fit with the
model given in Ref. \cite{Tanmoy2017pol} (\textit{Appendix}) and
the corresponding time constants are given in Table \ref{Tconsts}.}
\label{Dynamics}
\end{figure}

The amplitudes of the three spectral lines and their sum as a function
of the repolarizing laser pulse duration are shown in Fig. \ref{Dynamics}
for a single cycle of polarization transfer and repolarization. The
amplitudes of all three lines increase initially, indicating that
the electron spin polarization increases. The central line ($m_{I}=0$)
reaches its maximum after $\approx$500 ns and then starts to decrease,
whereas the outer lines ($m_{I}=-1$ and $+1$) continue to grow.
This indicates that the polarization of the nuclear spin decreases.
The sum of the amplitudes, after reaching its maximum value around
500 ns, stays roughly constant for the 532 nm illumination, whereas
for the 594 nm illumination it starts to decrease. This decay can
be fit to an exponentially decaying function with a time constant
of 20.4 $\mu$s. This decay reflects a decrease of the NV$^{-}$ population.
Its time constant is more than an order of magnitude longer than the
time needed to repolarize the center and hence does not cause significant
loss of signal. The rate constants for the polarization and depolarization
can be obtained by fitting the data to the model given in Ref. \cite{Tanmoy2017pol}
(\textit{Appendix}). The time constants for the polarization of the
electron spin and the depolarization of the $^{14}$N nuclear spin
for different wavelengths are given in Table \ref{Tconsts}. The polarization
rates for the wavelengths 532 and 594 nm are very similar but faster
compared to the one with 520 nm illumination. However the depolarization
rate is significantly slower for the 594 nm illumination compared
to the 520 and 532 nm ones. This implies that one should be able to
reach higher nuclear spin polarization with 594 nm light and it explains
the results of Fig. \ref{Spectra}.

\begin{table}
\begin{tabular*}{9cm}{@{\extracolsep{\fill}}|>{\centering}p{1.8cm}|>{\centering}p{2cm}|>{\centering}p{2.2cm}|>{\centering}p{2cm}|}
\hline 
Wavelength (nm) & Electron spin

polarization time constant (ns) & Nuclear spin depolarization time constant ($\mu$s) & Decay time of NV\textsuperscript{-} population ($\mu$s)\tabularnewline
\hline 
\hline 
520 & 170 ($\pm$21) & 6.4 ($\pm$1.1) & NA\tabularnewline
\hline 
532 & 101 ($\pm$16) & 8.4 ($\pm$2.5) & NA\tabularnewline
\hline 
594 & 110 ($\pm$22) & 16.6 ($\pm$4.8) & 20.4 ($\pm$1.0)\tabularnewline
\hline 
\end{tabular*}

\caption{Electron spin polarization and $^{14}$N nuclear spin depolarization
time constants for different wavelengths . The time constant corresponding
to the decay of the NV$^{-}$ population for the 594 nm illumination
is also given. The laser powers for 520 nm and 532 nm are $\approx$110
and 90 $\mu$W respectively. They are chosen such that the fluorescence
count rate is half of the saturation value. The laser power for 594
nm is $\approx$80 $\mu$W.}

\label{Tconsts}
\end{table}

\begin{figure}
\includegraphics[width=8cm]{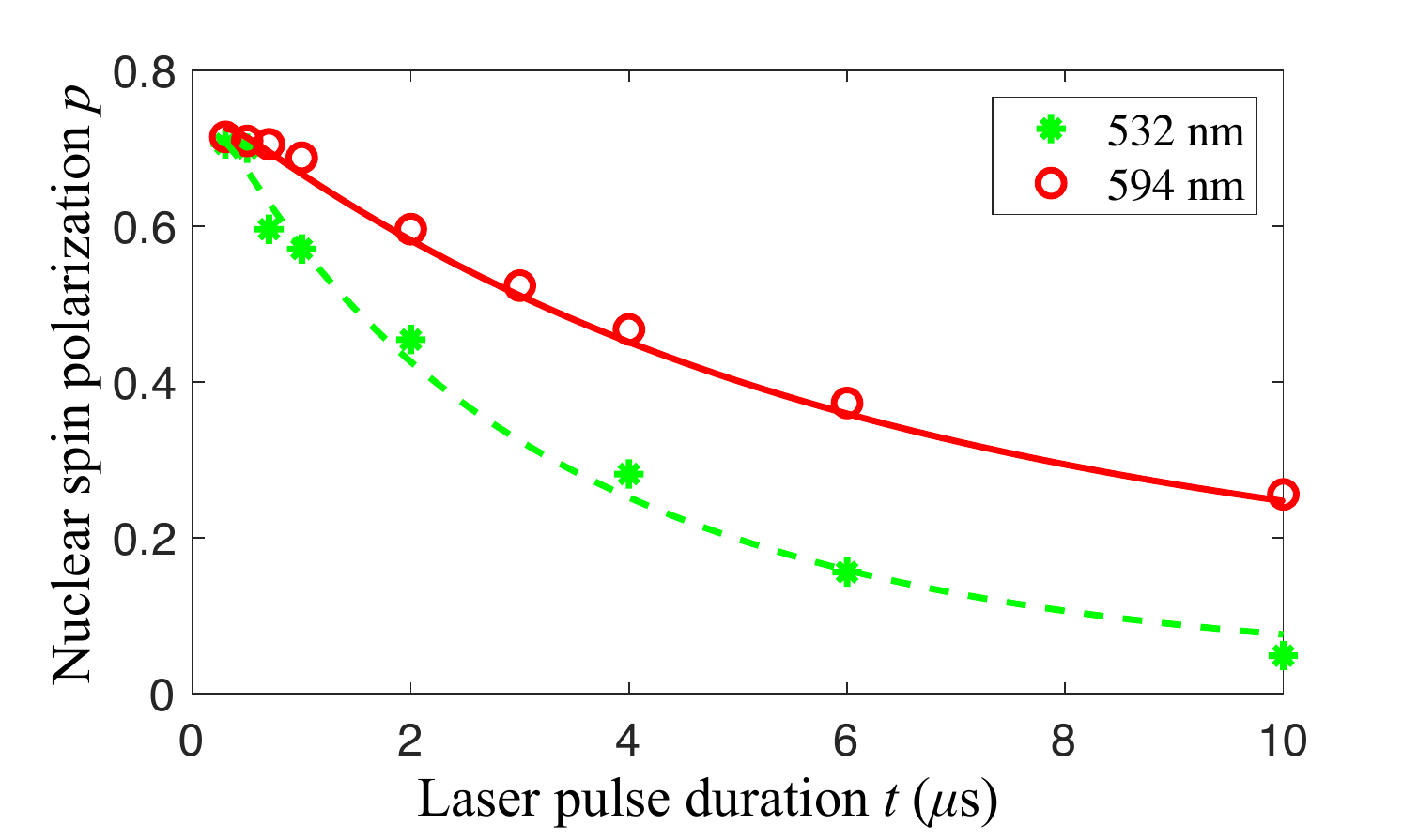}

\caption{Nuclear spin polarization as a function of laser pulse duration for
532 and 594 nm illumination. Circles and asterisks represent the experimental
data, solid and dashed curves are the corresponding fits according
to the model given in \textit{Appendix}.}

\label{Pol}
\end{figure}

The nuclear spin polarization measured from the data of Fig. \ref{Dynamics}
is shown in Fig. \ref{Pol} as a function of the laser pulse duration.
It clearly shows that the nuclear spin depolarization is slower for
594 nm illumination than for 532 nm.

\begin{figure}
\includegraphics[width=8cm]{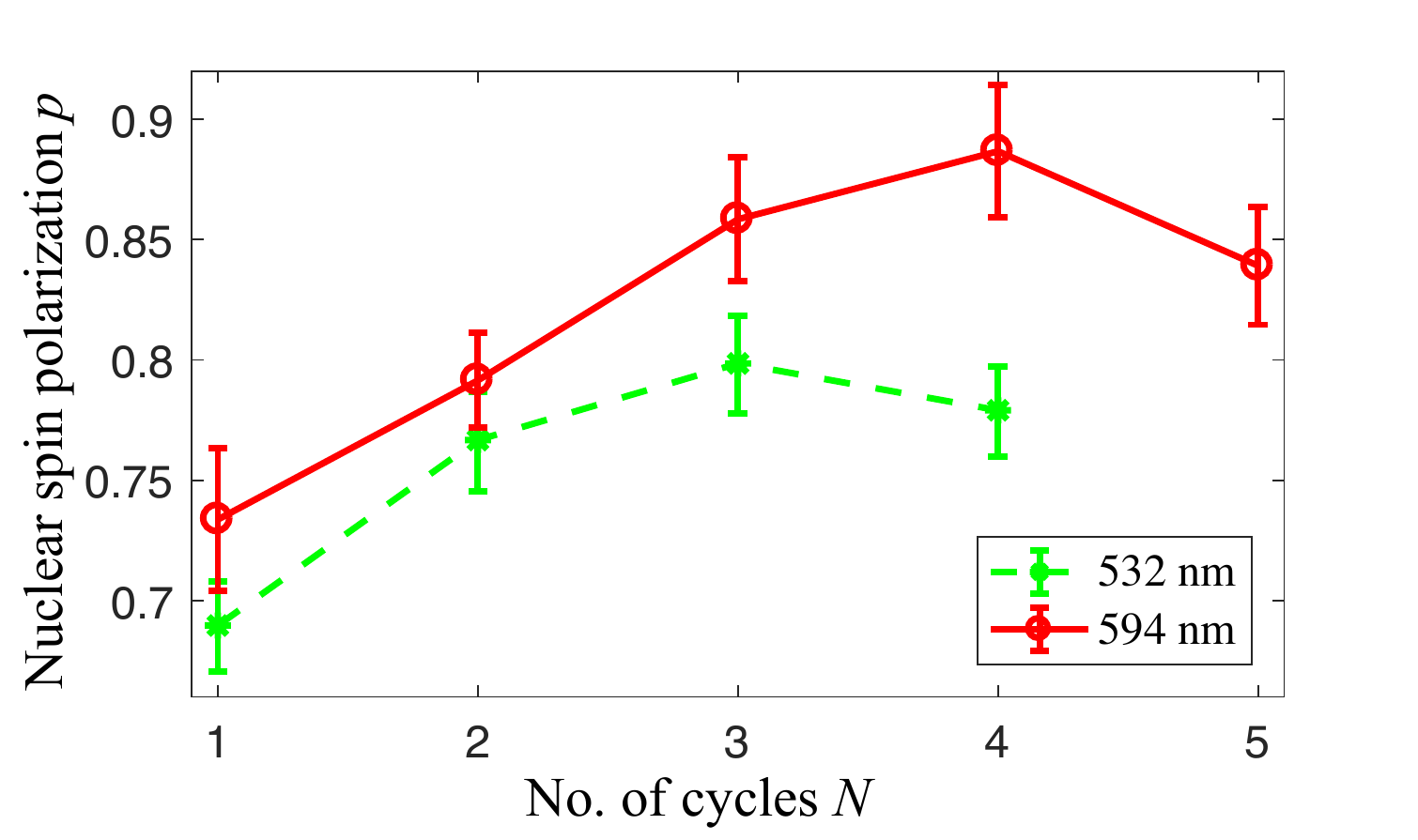}

\caption{Nuclear spin polarization versus $N$, the number of polarization
transfer and repolarization cycles. The duration of the repolarizing
laser pulse is 700 (500) ns for 594 (532) nm illumination.}

\label{PolvsCycles}
\end{figure}

Even higher polarizations can be achieved by iterating the transfer-repolarization
cycle, as indicated in Fig. \ref{Pulseseq}. Fig. \ref{PolvsCycles}
shows the nuclear spin polarization as a function of the number of
cycles for both the 532 and 594 nm repolarizing illuminations. The
polarization increases with the number of cycles for both the cases
and reaches its maximum value after 4 (3) cycles for the 594 (532)
nm illumination. The low value of the nuclear spin polarization for
N=1 could be attributed to the imperfect MW and RF pulses and other
experimental non-idealities which leave significant population in
the $m_{I}=-1$ and $+1$ states. By iterating the transfer-repolarization
cycle this population can be repumped into the $m_{I}=0$ state and
hence improves the nuclear spin polarization.

Orange light illumination improves the $^{14}$N nuclear spin polarization
but it also leads to loss of total signal. Here, we analyze the data
given in Fig. \ref{Spectra} to understand signal loss and gain in
polarization. From this data, the signal of the electron spin, in
the order of nuclear spin states $m_{I}=-1,\,0,\,+1$, can be written
as {[}9.96 94.61 8.49{]} (for green repolarizing pulses) and {[}3.38
95.73 4.61{]} (for orange repolarizing pulses). These signals are proportional
to the populations of the corresponding nuclear spin states. So, the
total signal loss is $\sim$ 8 \% and the gain in polarization is
$\sim$ 13 \%. However, the signal of the $m_{I}=0$ state is roughly
equal in both cases and the loss of signal is only reflected in the
$m_{I}=-1$ and $+1$ states. This implies, the signal that is lost
is originally noise in the present scenario (of initializing quantum
registers). In any case, the loss of signal can be compensated by
increasing the averaging time by 17\%.

\section{Discussion and Conclusion}

Initializing single nuclear spins coupled to electron spins in solid
state materials is an important prerequisite for spin based hybrid
quantum information processing and other applications like sensing.
The approach described here, using a combination of laser (green,
532 or 520 nm), MW, and RF pulses allows one to initialize nuclear
spins coupled to NV centers in diamond at arbitrary magnetic fields.
By judicious use of 532 and 594 nm (orange) laser pulses, we can achieve
higher nuclear spin polarization (89.0 \%) than by using only 532
nm laser pulses (76.3 \%). This polarization can be improved further
by improving the fidelities of the transition selective MW and RF
pulses. We have also studied the dynamics of electron spin polarization
and nuclear spin depolarization under different wavelengths of laser
illumination. We found that the electron spin polarization rates are
similar for the laser wavelengths 532 and 594 nm and the nuclear spin
depolarization rate under 594 nm illumination is significantly smaller
than with 532 nm. We believe the reason for different nuclear spin
depolarization rates is related to the photo-induced ionization of
the NV center. It is known that green light can ionize NV$^{-}$ into
NV$^{0}$ and vice-versa, and orange light also can ionize NV$^{-}$
into NV$^{0}$, but the reverse processis less probable \cite{Nabeel_Ion}.
This combined with the previous observation \cite{Nabeel_Ion}, that
a single NV center in the NV$^{0}$ state yields very little fluorescence
under green light irradiation can explain the observed slower nuclear
spin depolarization under 594 nm illumination.

The method of polarizing nuclear spins by using a combination of MW,
RF, and laser pulses has been previously applied to polarize $^{13}$C
nuclear spins as well \cite{Shim2013,DuChopped2019}. Here, we note
that the achievable polarization by this method might also depend
on the form of the hyperfine interaction. For the $^{14}$N nuclear
spin, the hyperfine tensor contains no off diagonal elements \cite{Mansion14Nhyperfine,Felton2009,PaolaPRB2015}
and the effect of the transverse (diagonal) components of the hyperfine
tensor, which are off-diagonal to the Hamiltonian, can be neglected
to an approximation. This implies that the nuclear spin eigenstates
are approximately the Zeeman states. However, for $^{13}$C nuclear
spins, the hyperfine tensor in general contains off-diagonal elements,
in particular terms that commute with the $z$-component of the electron
spin angular momentum but not with the $z$-component of the nuclear
spin angular momentum \cite{Jacques13Chyperfine,ShimArXiv,Koti2016PRB}.
This leads to mixing of nuclear spin Zeeman states within the $m_{S}=-1$
and $+1$ subspaces and for the $m_{S}=0$ subspace, Zeeman states
are still approximately the eigenstates if the magnetic field is aligned
with the NV axis. In any case, the use of green and orange light illumination
should lead to higher nuclear spin polarization compared to green
light illumination alone even for $^{13}$C nuclear spins.

One disadvantage of the 594 nm excitation is the loss of NV$^{-}$
population, which corresponds to a reduction of the total signal.
The loss of signal can be compensated by taking more averages, but
the purity of a quantum state cannot be improved in a similar way.
High purity quantum states are important for many applications. For
example, the amount of entanglement that can be generated in a system
depends ultimately on the purity of its initial state \cite{MortonEnt2011}.
High purity quantum states are also desirable for experiments involving
fundamental tests of quantum mechanics such as temporal Bell inequality
\cite{WaldherrLG2011}.

\section{Acknowledgments}

This work was supported by the DFG through Grant No. Su 192/28-1. This project has received funding from the
European Union's Horizon 2020 research and innovation
programme under grant agreement No 828946. The publication reflects
the opinion of the authors; the agency and the commission may not
be held responsible for the information contained in it.

\section*{Appendix}
\begin{center}
\textbf{Rate equation model}
\par\end{center}

The rate equation model for the population dynamics of an NV center
under laser illumination can be written as

\begin{equation}
\frac{d}{dt}\vec{P}=\left(M(k_{S},k_{I})-k_{p}\right)\vec{P},
\end{equation}
where $k_{S}$, $k_{I}$, and $k_{p}$ represent the rate constants
for the electron spin polarization, nuclear spin depolarization, and
the decay of the NV\textsuperscript{-} population respectively. $\vec{P}$
and $M(k_{S},k_{I})$ are the population vector and transition matrix
respectively, defined as

\begin{multline*}
\vec{P}=\left(P_{\left|m_{S},\,m_{I}\right\rangle =\left|0,\,+1\right\rangle },\:P_{\left|0,\,-1\right\rangle },\:P_{\left|0,\,0\right\rangle },\:P_{\left|-1,\,-1\right\rangle },\right.\\
\left.\:P_{\left|-1,\,+1\right\rangle },\:P_{\left|-1,\,0\right\rangle },\:P_{\left|+1,\,+1\right\rangle },\:P_{\left|+1,\,-1\right\rangle },\:P_{\left|+1,\,0\right\rangle }\right),
\end{multline*}

\begin{multline*}
M(k_{S},k_{I})=\\
\left[\begin{array}{ccccccccc}
-2k_{I} & k_{I} & k_{I} & 0 & k_{S} & 0 & k_{S} & 0 & 0\\
k_{I} & -2k_{I} & k_{I} & k_{S} & 0 & 0 & 0 & k_{S} & 0\\
k_{I} & k_{I} & -2k_{I} & 0 & 0 & k_{S} & 0 & 0 & k_{S}\\
0 & 0 & 0 & -k_{S} & 0 & 0 & 0 & 0 & 0\\
0 & 0 & 0 & 0 & -k_{S} & 0 & 0 & 0 & 0\\
0 & 0 & 0 & 0 & 0 & -k_{S} & 0 & 0 & 0\\
0 & 0 & 0 & 0 & 0 &  & -k_{S} & 0 & 0\\
0 & 0 & 0 & 0 & 0 & 0 & 0 & -k_{S} & 0\\
0 & 0 & 0 & 0 & 0 & 0 & 0 & 0 & -k_{S}
\end{array}\right].
\end{multline*}

The population vector just before the repolarizing laser pulse can
be written as $\vec{P}=\frac{1}{3}\left(0,\,0,\,1,0,\,0,\,1,0,\,0,\,1\right)$.
Using this vector as the initial state, the solution to the rate equation
model can be obtained as
\begin{multline*}
\vec{P}=\frac{1}{3}e^{-k_{p}t}\left(1-\frac{2k_{I}}{(3k_{I}-k_{S})}e^{-k_{S}t}-\frac{(k_{I}-k_{S})}{(3k_{I}-k_{S})}e^{-3k_{I}t},\right.\\
1-\frac{2k_{I}}{(3k_{I}-k_{S})}e^{-k_{S}t}-\frac{(k_{I}-k_{S})}{(3k_{I}-k_{S})}e^{-3k_{I}t},\\
1-\frac{2(k_{I}-k_{S})}{(3k_{I}-k_{S})}e^{-k_{S}t}+\frac{2(k_{I}-k_{S})}{(3k_{I}-k_{S})}e^{-3k_{I}t},\\
\left.0,\,0,\,e^{-k_{S}t},\,0,\,0,\,e^{-k_{S}t}\right).
\end{multline*}
This vector represents the population dynamics under the repolarizing
laser pulse.
\begin{center}
\par\end{center}

\bibliography{bibNV1}

\begin{thebibliography}{34}%
\makeatletter
\providecommand \@ifxundefined [1]{%
 \@ifx{#1\undefined}
}%
\providecommand \@ifnum [1]{%
 \ifnum #1\expandafter \@firstoftwo
 \else \expandafter \@secondoftwo
 \fi
}%
\providecommand \@ifx [1]{%
 \ifx #1\expandafter \@firstoftwo
 \else \expandafter \@secondoftwo
 \fi
}%
\providecommand \natexlab [1]{#1}%
\providecommand \enquote  [1]{``#1''}%
\providecommand \bibnamefont  [1]{#1}%
\providecommand \bibfnamefont [1]{#1}%
\providecommand \citenamefont [1]{#1}%
\providecommand \href@noop [0]{\@secondoftwo}%
\providecommand \href [0]{\begingroup \@sanitize@url \@href}%
\providecommand \@href[1]{\@@startlink{#1}\@@href}%
\providecommand \@@href[1]{\endgroup#1\@@endlink}%
\providecommand \@sanitize@url [0]{\catcode `\\12\catcode `\$12\catcode
  `\&12\catcode `\#12\catcode `\^12\catcode `\_12\catcode `\%12\relax}%
\providecommand \@@startlink[1]{}%
\providecommand \@@endlink[0]{}%
\providecommand \url  [0]{\begingroup\@sanitize@url \@url }%
\providecommand \@url [1]{\endgroup\@href {#1}{\urlprefix }}%
\providecommand \urlprefix  [0]{URL }%
\providecommand \Eprint [0]{\href }%
\providecommand \doibase [0]{http://dx.doi.org/}%
\providecommand \selectlanguage [0]{\@gobble}%
\providecommand \bibinfo  [0]{\@secondoftwo}%
\providecommand \bibfield  [0]{\@secondoftwo}%
\providecommand \translation [1]{[#1]}%
\providecommand \BibitemOpen [0]{}%
\providecommand \bibitemStop [0]{}%
\providecommand \bibitemNoStop [0]{.\EOS\space}%
\providecommand \EOS [0]{\spacefactor3000\relax}%
\providecommand \BibitemShut  [1]{\csname bibitem#1\endcsname}%
\let\auto@bib@innerbib\@empty
\bibitem [{\citenamefont {Childress}\ \emph {et~al.}(2006)\citenamefont
  {Childress}, \citenamefont {Gurudev~Dutt}, \citenamefont {Taylor},
  \citenamefont {Zibrov}, \citenamefont {Jelezko}, \citenamefont {Wrachtrup},
  \citenamefont {Hemmer},\ and\ \citenamefont {Lukin}}]{Chil2006Sci}%
  \BibitemOpen
  \bibfield  {author} {\bibinfo {author} {\bibfnamefont {L.}~\bibnamefont
  {Childress}}, \bibinfo {author} {\bibfnamefont {M.~V.}\ \bibnamefont
  {Gurudev~Dutt}}, \bibinfo {author} {\bibfnamefont {J.~M.}\ \bibnamefont
  {Taylor}}, \bibinfo {author} {\bibfnamefont {A.~S.}\ \bibnamefont {Zibrov}},
  \bibinfo {author} {\bibfnamefont {F.}~\bibnamefont {Jelezko}}, \bibinfo
  {author} {\bibfnamefont {J.}~\bibnamefont {Wrachtrup}}, \bibinfo {author}
  {\bibfnamefont {P.~R.}\ \bibnamefont {Hemmer}}, \ and\ \bibinfo {author}
  {\bibfnamefont {M.~D.}\ \bibnamefont {Lukin}},\ }\href {\doibase
  10.1126/science.1131871} {\bibfield  {journal} {\bibinfo  {journal}
  {Science}\ }\textbf {\bibinfo {volume} {314}},\ \bibinfo {pages} {281}
  (\bibinfo {year} {2006})}\BibitemShut {NoStop}%
\bibitem [{\citenamefont {Dutt}\ \emph {et~al.}(2007)\citenamefont {Dutt},
  \citenamefont {Childress}, \citenamefont {Jiang}, \citenamefont {Togan},
  \citenamefont {Maze}, \citenamefont {Jelezko}, \citenamefont {Zibrov},
  \citenamefont {Hemmer},\ and\ \citenamefont {Lukin}}]{Lukin2007Sci}%
  \BibitemOpen
  \bibfield  {author} {\bibinfo {author} {\bibfnamefont {M.~V.~G.}\
  \bibnamefont {Dutt}}, \bibinfo {author} {\bibfnamefont {L.}~\bibnamefont
  {Childress}}, \bibinfo {author} {\bibfnamefont {L.}~\bibnamefont {Jiang}},
  \bibinfo {author} {\bibfnamefont {E.}~\bibnamefont {Togan}}, \bibinfo
  {author} {\bibfnamefont {J.}~\bibnamefont {Maze}}, \bibinfo {author}
  {\bibfnamefont {F.}~\bibnamefont {Jelezko}}, \bibinfo {author} {\bibfnamefont
  {A.~S.}\ \bibnamefont {Zibrov}}, \bibinfo {author} {\bibfnamefont {P.~R.}\
  \bibnamefont {Hemmer}}, \ and\ \bibinfo {author} {\bibfnamefont {M.~D.}\
  \bibnamefont {Lukin}},\ }\href {\doibase 10.1126/science.1139831} {\bibfield
  {journal} {\bibinfo  {journal} {Science}\ }\textbf {\bibinfo {volume}
  {316}},\ \bibinfo {pages} {1312} (\bibinfo {year} {2007})}\BibitemShut
  {NoStop}%
\bibitem [{\citenamefont {Staudacher}\ \emph {et~al.}(2013)\citenamefont
  {Staudacher}, \citenamefont {Shi}, \citenamefont {Pezzagna}, \citenamefont
  {Meijer}, \citenamefont {Du}, \citenamefont {Meriles}, \citenamefont
  {Reinhard},\ and\ \citenamefont {Wrachtrup}}]{JW2013Sci}%
  \BibitemOpen
  \bibfield  {author} {\bibinfo {author} {\bibfnamefont {T.}~\bibnamefont
  {Staudacher}}, \bibinfo {author} {\bibfnamefont {F.}~\bibnamefont {Shi}},
  \bibinfo {author} {\bibfnamefont {S.}~\bibnamefont {Pezzagna}}, \bibinfo
  {author} {\bibfnamefont {J.}~\bibnamefont {Meijer}}, \bibinfo {author}
  {\bibfnamefont {J.}~\bibnamefont {Du}}, \bibinfo {author} {\bibfnamefont
  {C.~A.}\ \bibnamefont {Meriles}}, \bibinfo {author} {\bibfnamefont
  {F.}~\bibnamefont {Reinhard}}, \ and\ \bibinfo {author} {\bibfnamefont
  {J.}~\bibnamefont {Wrachtrup}},\ }\href {\doibase 10.1126/science.1231675}
  {\bibfield  {journal} {\bibinfo  {journal} {Science}\ }\textbf {\bibinfo
  {volume} {339}},\ \bibinfo {pages} {561} (\bibinfo {year}
  {2013})}\BibitemShut {NoStop}%
\bibitem [{\citenamefont {Waldherr}\ \emph {et~al.}(2014)\citenamefont
  {Waldherr}, \citenamefont {Wang}, \citenamefont {Zaiser}, \citenamefont
  {Jamali}, \citenamefont {Schulte-Herbr{\"u}ggen}, \citenamefont {Abe},
  \citenamefont {Ohshima}, \citenamefont {Isoya}, \citenamefont {Du},
  \citenamefont {Neumann} \emph {et~al.}}]{JW2014Nat}%
  \BibitemOpen
  \bibfield  {author} {\bibinfo {author} {\bibfnamefont {G.}~\bibnamefont
  {Waldherr}}, \bibinfo {author} {\bibfnamefont {Y.}~\bibnamefont {Wang}},
  \bibinfo {author} {\bibfnamefont {S.}~\bibnamefont {Zaiser}}, \bibinfo
  {author} {\bibfnamefont {M.}~\bibnamefont {Jamali}}, \bibinfo {author}
  {\bibfnamefont {T.}~\bibnamefont {Schulte-Herbr{\"u}ggen}}, \bibinfo {author}
  {\bibfnamefont {H.}~\bibnamefont {Abe}}, \bibinfo {author} {\bibfnamefont
  {T.}~\bibnamefont {Ohshima}}, \bibinfo {author} {\bibfnamefont
  {J.}~\bibnamefont {Isoya}}, \bibinfo {author} {\bibfnamefont
  {J.}~\bibnamefont {Du}}, \bibinfo {author} {\bibfnamefont {P.}~\bibnamefont
  {Neumann}},  \emph {et~al.},\ }\href {\doibase 10.1038/nature12919}
  {\bibfield  {journal} {\bibinfo  {journal} {Nature}\ }\textbf {\bibinfo
  {volume} {506}},\ \bibinfo {pages} {204} (\bibinfo {year}
  {2014})}\BibitemShut {NoStop}%
\bibitem [{\citenamefont {Doherty}\ \emph {et~al.}(2013)\citenamefont
  {Doherty}, \citenamefont {Manson}, \citenamefont {Delaney}, \citenamefont
  {Jelezko}, \citenamefont {Wrachtrup},\ and\ \citenamefont
  {Hollenberg}}]{RevDoherty}%
  \BibitemOpen
  \bibfield  {author} {\bibinfo {author} {\bibfnamefont {M.~W.}\ \bibnamefont
  {Doherty}}, \bibinfo {author} {\bibfnamefont {N.~B.}\ \bibnamefont {Manson}},
  \bibinfo {author} {\bibfnamefont {P.}~\bibnamefont {Delaney}}, \bibinfo
  {author} {\bibfnamefont {F.}~\bibnamefont {Jelezko}}, \bibinfo {author}
  {\bibfnamefont {J.}~\bibnamefont {Wrachtrup}}, \ and\ \bibinfo {author}
  {\bibfnamefont {L.~C.}\ \bibnamefont {Hollenberg}},\ }\href {\doibase
  http://dx.doi.org/10.1016/j.physrep.2013.02.001} {\bibfield  {journal}
  {\bibinfo  {journal} {Physics Reports}\ }\textbf {\bibinfo {volume} {528}},\
  \bibinfo {pages} {1 } (\bibinfo {year} {2013})}\BibitemShut {NoStop}%
\bibitem [{\citenamefont {Childress}\ and\ \citenamefont
  {Hanson}(2013)}]{RevChildress}%
  \BibitemOpen
  \bibfield  {author} {\bibinfo {author} {\bibfnamefont {L.}~\bibnamefont
  {Childress}}\ and\ \bibinfo {author} {\bibfnamefont {R.}~\bibnamefont
  {Hanson}},\ }\href {\doibase 10.1557/mrs.2013.20} {\bibfield  {journal}
  {\bibinfo  {journal} {MRS Bulletin}\ }\textbf {\bibinfo {volume} {38}},\
  \bibinfo {pages} {134} (\bibinfo {year} {2013})}\BibitemShut {NoStop}%
\bibitem [{\citenamefont {Hong}\ \emph {et~al.}(2013)\citenamefont {Hong},
  \citenamefont {Grinolds}, \citenamefont {Pham}, \citenamefont {Sage},
  \citenamefont {Luan}, \citenamefont {Walsworth},\ and\ \citenamefont
  {Yacoby}}]{RevWalsworth}%
  \BibitemOpen
  \bibfield  {author} {\bibinfo {author} {\bibfnamefont {S.}~\bibnamefont
  {Hong}}, \bibinfo {author} {\bibfnamefont {M.~S.}\ \bibnamefont {Grinolds}},
  \bibinfo {author} {\bibfnamefont {L.~M.}\ \bibnamefont {Pham}}, \bibinfo
  {author} {\bibfnamefont {D.~L.}\ \bibnamefont {Sage}}, \bibinfo {author}
  {\bibfnamefont {L.}~\bibnamefont {Luan}}, \bibinfo {author} {\bibfnamefont
  {R.~L.}\ \bibnamefont {Walsworth}}, \ and\ \bibinfo {author} {\bibfnamefont
  {A.}~\bibnamefont {Yacoby}},\ }\href {\doibase 10.1557/mrs.2013.23}
  {\bibfield  {journal} {\bibinfo  {journal} {MRS Bulletin}\ }\textbf {\bibinfo
  {volume} {38}},\ \bibinfo {pages} {155} (\bibinfo {year} {2013})}\BibitemShut
  {NoStop}%
\bibitem [{\citenamefont {Schirhagl}\ \emph {et~al.}(2014)\citenamefont
  {Schirhagl}, \citenamefont {Chang}, \citenamefont {Loretz},\ and\
  \citenamefont {Degen}}]{RevDegen}%
  \BibitemOpen
  \bibfield  {author} {\bibinfo {author} {\bibfnamefont {R.}~\bibnamefont
  {Schirhagl}}, \bibinfo {author} {\bibfnamefont {K.}~\bibnamefont {Chang}},
  \bibinfo {author} {\bibfnamefont {M.}~\bibnamefont {Loretz}}, \ and\ \bibinfo
  {author} {\bibfnamefont {C.~L.}\ \bibnamefont {Degen}},\ }\href {\doibase
  10.1146/annurev-physchem-040513-103659} {\bibfield  {journal} {\bibinfo
  {journal} {Annual Review of Physical Chemistry}\ }\textbf {\bibinfo {volume}
  {65}},\ \bibinfo {pages} {83} (\bibinfo {year} {2014})}\BibitemShut {NoStop}%
\bibitem [{\citenamefont {Neumann}\ \emph {et~al.}(2008)\citenamefont
  {Neumann}, \citenamefont {Mizuochi}, \citenamefont {Rempp}, \citenamefont
  {Hemmer}, \citenamefont {Watanabe}, \citenamefont {Yamasaki}, \citenamefont
  {Jacques}, \citenamefont {Gaebel}, \citenamefont {Jelezko},\ and\
  \citenamefont {Wrachtrup}}]{JW2008Sci}%
  \BibitemOpen
  \bibfield  {author} {\bibinfo {author} {\bibfnamefont {P.}~\bibnamefont
  {Neumann}}, \bibinfo {author} {\bibfnamefont {N.}~\bibnamefont {Mizuochi}},
  \bibinfo {author} {\bibfnamefont {F.}~\bibnamefont {Rempp}}, \bibinfo
  {author} {\bibfnamefont {P.}~\bibnamefont {Hemmer}}, \bibinfo {author}
  {\bibfnamefont {H.}~\bibnamefont {Watanabe}}, \bibinfo {author}
  {\bibfnamefont {S.}~\bibnamefont {Yamasaki}}, \bibinfo {author}
  {\bibfnamefont {V.}~\bibnamefont {Jacques}}, \bibinfo {author} {\bibfnamefont
  {T.}~\bibnamefont {Gaebel}}, \bibinfo {author} {\bibfnamefont
  {F.}~\bibnamefont {Jelezko}}, \ and\ \bibinfo {author} {\bibfnamefont
  {J.}~\bibnamefont {Wrachtrup}},\ }\href {\doibase 10.1126/science.1157233}
  {\bibfield  {journal} {\bibinfo  {journal} {Science}\ }\textbf {\bibinfo
  {volume} {320}},\ \bibinfo {pages} {1326} (\bibinfo {year}
  {2008})}\BibitemShut {NoStop}%
\bibitem [{\citenamefont {Taminiau}\ \emph {et~al.}(2014)\citenamefont
  {Taminiau}, \citenamefont {Cramer}, \citenamefont {van~der Sar},
  \citenamefont {Dobrovitski},\ and\ \citenamefont {Hanson}}]{Han2014NatNano}%
  \BibitemOpen
  \bibfield  {author} {\bibinfo {author} {\bibfnamefont {T.~H.}\ \bibnamefont
  {Taminiau}}, \bibinfo {author} {\bibfnamefont {J.}~\bibnamefont {Cramer}},
  \bibinfo {author} {\bibfnamefont {T.}~\bibnamefont {van~der Sar}}, \bibinfo
  {author} {\bibfnamefont {V.~V.}\ \bibnamefont {Dobrovitski}}, \ and\ \bibinfo
  {author} {\bibfnamefont {R.}~\bibnamefont {Hanson}},\ }\href {\doibase
  10.1038/nnano.2014.2} {\bibfield  {journal} {\bibinfo  {journal} {Nature
  nanotechnology}\ }\textbf {\bibinfo {volume} {9}},\ \bibinfo {pages} {171}
  (\bibinfo {year} {2014})}\BibitemShut {NoStop}%
\bibitem [{\citenamefont {Zhang}\ and\ \citenamefont
  {Suter}(2015)}]{ZhangPRL2015}%
  \BibitemOpen
  \bibfield  {author} {\bibinfo {author} {\bibfnamefont {J.}~\bibnamefont
  {Zhang}}\ and\ \bibinfo {author} {\bibfnamefont {D.}~\bibnamefont {Suter}},\
  }\href {\doibase 10.1103/PhysRevLett.115.110502} {\bibfield  {journal}
  {\bibinfo  {journal} {Phys. Rev. Lett.}\ }\textbf {\bibinfo {volume} {115}},\
  \bibinfo {pages} {110502} (\bibinfo {year} {2015})}\BibitemShut {NoStop}%
\bibitem [{\citenamefont {Fuchs}\ \emph {et~al.}(2011)\citenamefont {Fuchs},
  \citenamefont {Burkard}, \citenamefont {Klimov},\ and\ \citenamefont
  {Awschalom}}]{Awsch2011NatPhy}%
  \BibitemOpen
  \bibfield  {author} {\bibinfo {author} {\bibfnamefont {G.~D.}\ \bibnamefont
  {Fuchs}}, \bibinfo {author} {\bibfnamefont {G.}~\bibnamefont {Burkard}},
  \bibinfo {author} {\bibfnamefont {P.~V.}\ \bibnamefont {Klimov}}, \ and\
  \bibinfo {author} {\bibfnamefont {D.~D.}\ \bibnamefont {Awschalom}},\ }\href
  {\doibase 10.1038/nphys2026} {\bibfield  {journal} {\bibinfo  {journal}
  {Nature Physics}\ }\textbf {\bibinfo {volume} {7}},\ \bibinfo {pages} {789}
  (\bibinfo {year} {2011})}\BibitemShut {NoStop}%
\bibitem [{\citenamefont {Maurer}\ \emph {et~al.}(2012)\citenamefont {Maurer},
  \citenamefont {Kucsko}, \citenamefont {Latta}, \citenamefont {Jiang},
  \citenamefont {Yao}, \citenamefont {Bennett}, \citenamefont {Pastawski},
  \citenamefont {Hunger}, \citenamefont {Chisholm}, \citenamefont {Markham},
  \citenamefont {Twitchen}, \citenamefont {Cirac},\ and\ \citenamefont
  {Lukin}}]{Lukin2012Sci}%
  \BibitemOpen
  \bibfield  {author} {\bibinfo {author} {\bibfnamefont {P.~C.}\ \bibnamefont
  {Maurer}}, \bibinfo {author} {\bibfnamefont {G.}~\bibnamefont {Kucsko}},
  \bibinfo {author} {\bibfnamefont {C.}~\bibnamefont {Latta}}, \bibinfo
  {author} {\bibfnamefont {L.}~\bibnamefont {Jiang}}, \bibinfo {author}
  {\bibfnamefont {N.~Y.}\ \bibnamefont {Yao}}, \bibinfo {author} {\bibfnamefont
  {S.~D.}\ \bibnamefont {Bennett}}, \bibinfo {author} {\bibfnamefont
  {F.}~\bibnamefont {Pastawski}}, \bibinfo {author} {\bibfnamefont
  {D.}~\bibnamefont {Hunger}}, \bibinfo {author} {\bibfnamefont
  {N.}~\bibnamefont {Chisholm}}, \bibinfo {author} {\bibfnamefont
  {M.}~\bibnamefont {Markham}}, \bibinfo {author} {\bibfnamefont {D.~J.}\
  \bibnamefont {Twitchen}}, \bibinfo {author} {\bibfnamefont {J.~I.}\
  \bibnamefont {Cirac}}, \ and\ \bibinfo {author} {\bibfnamefont {M.~D.}\
  \bibnamefont {Lukin}},\ }\href {\doibase 10.1126/science.1220513} {\bibfield
  {journal} {\bibinfo  {journal} {Science}\ }\textbf {\bibinfo {volume}
  {336}},\ \bibinfo {pages} {1283} (\bibinfo {year} {2012})}\BibitemShut
  {NoStop}%
\bibitem [{\citenamefont {Shim}\ \emph
  {et~al.}(2013{\natexlab{a}})\citenamefont {Shim}, \citenamefont {Niemeyer},
  \citenamefont {Zhang},\ and\ \citenamefont {Suter}}]{Shim2013}%
  \BibitemOpen
  \bibfield  {author} {\bibinfo {author} {\bibfnamefont {J.~H.}\ \bibnamefont
  {Shim}}, \bibinfo {author} {\bibfnamefont {I.}~\bibnamefont {Niemeyer}},
  \bibinfo {author} {\bibfnamefont {J.}~\bibnamefont {Zhang}}, \ and\ \bibinfo
  {author} {\bibfnamefont {D.}~\bibnamefont {Suter}},\ }\href {\doibase
  10.1103/PhysRevA.87.012301} {\bibfield  {journal} {\bibinfo  {journal} {Phys.
  Rev. A}\ }\textbf {\bibinfo {volume} {87}},\ \bibinfo {pages} {012301}
  (\bibinfo {year} {2013}{\natexlab{a}})}\BibitemShut {NoStop}%
\bibitem [{\citenamefont {Fischer}\ \emph {et~al.}(2013)\citenamefont
  {Fischer}, \citenamefont {Bretschneider}, \citenamefont {London},
  \citenamefont {Budker}, \citenamefont {Gershoni},\ and\ \citenamefont
  {Frydman}}]{Frydman2013hyperpol}%
  \BibitemOpen
  \bibfield  {author} {\bibinfo {author} {\bibfnamefont {R.}~\bibnamefont
  {Fischer}}, \bibinfo {author} {\bibfnamefont {C.~O.}\ \bibnamefont
  {Bretschneider}}, \bibinfo {author} {\bibfnamefont {P.}~\bibnamefont
  {London}}, \bibinfo {author} {\bibfnamefont {D.}~\bibnamefont {Budker}},
  \bibinfo {author} {\bibfnamefont {D.}~\bibnamefont {Gershoni}}, \ and\
  \bibinfo {author} {\bibfnamefont {L.}~\bibnamefont {Frydman}},\ }\href
  {\doibase 10.1103/PhysRevLett.111.057601} {\bibfield  {journal} {\bibinfo
  {journal} {Phys. Rev. Lett.}\ }\textbf {\bibinfo {volume} {111}},\ \bibinfo
  {pages} {057601} (\bibinfo {year} {2013})}\BibitemShut {NoStop}%
\bibitem [{\citenamefont {{\'A}lvarez}\ \emph {et~al.}(2015)\citenamefont
  {{\'A}lvarez}, \citenamefont {Bretschneider}, \citenamefont {Fischer},
  \citenamefont {London}, \citenamefont {Kanda}, \citenamefont {Onoda},
  \citenamefont {Isoya}, \citenamefont {Gershoni},\ and\ \citenamefont
  {Frydman}}]{Alvarez2015}%
  \BibitemOpen
  \bibfield  {author} {\bibinfo {author} {\bibfnamefont {G.~A.}\ \bibnamefont
  {{\'A}lvarez}}, \bibinfo {author} {\bibfnamefont {C.~O.}\ \bibnamefont
  {Bretschneider}}, \bibinfo {author} {\bibfnamefont {R.}~\bibnamefont
  {Fischer}}, \bibinfo {author} {\bibfnamefont {P.}~\bibnamefont {London}},
  \bibinfo {author} {\bibfnamefont {H.}~\bibnamefont {Kanda}}, \bibinfo
  {author} {\bibfnamefont {S.}~\bibnamefont {Onoda}}, \bibinfo {author}
  {\bibfnamefont {J.}~\bibnamefont {Isoya}}, \bibinfo {author} {\bibfnamefont
  {D.}~\bibnamefont {Gershoni}}, \ and\ \bibinfo {author} {\bibfnamefont
  {L.}~\bibnamefont {Frydman}},\ }\href {\doibase 10.1038/ncomms9456}
  {\bibfield  {journal} {\bibinfo  {journal} {Nature communications}\ }\textbf
  {\bibinfo {volume} {6}},\ \bibinfo {pages} {8456} (\bibinfo {year}
  {2015})}\BibitemShut {NoStop}%
\bibitem [{\citenamefont {Pagliero}\ \emph {et~al.}(2018)\citenamefont
  {Pagliero}, \citenamefont {Rao}, \citenamefont {Zangara}, \citenamefont
  {Dhomkar}, \citenamefont {Wong}, \citenamefont {Abril}, \citenamefont
  {Aslam}, \citenamefont {Parker}, \citenamefont {King}, \citenamefont
  {Avalos}, \citenamefont {Ajoy}, \citenamefont {Wrachtrup}, \citenamefont
  {Pines},\ and\ \citenamefont {Meriles}}]{My_hyperpol2018}%
  \BibitemOpen
  \bibfield  {author} {\bibinfo {author} {\bibfnamefont {D.}~\bibnamefont
  {Pagliero}}, \bibinfo {author} {\bibfnamefont {K.~R.~K.}\ \bibnamefont
  {Rao}}, \bibinfo {author} {\bibfnamefont {P.~R.}\ \bibnamefont {Zangara}},
  \bibinfo {author} {\bibfnamefont {S.}~\bibnamefont {Dhomkar}}, \bibinfo
  {author} {\bibfnamefont {H.~H.}\ \bibnamefont {Wong}}, \bibinfo {author}
  {\bibfnamefont {A.}~\bibnamefont {Abril}}, \bibinfo {author} {\bibfnamefont
  {N.}~\bibnamefont {Aslam}}, \bibinfo {author} {\bibfnamefont
  {A.}~\bibnamefont {Parker}}, \bibinfo {author} {\bibfnamefont
  {J.}~\bibnamefont {King}}, \bibinfo {author} {\bibfnamefont {C.~E.}\
  \bibnamefont {Avalos}}, \bibinfo {author} {\bibfnamefont {A.}~\bibnamefont
  {Ajoy}}, \bibinfo {author} {\bibfnamefont {J.}~\bibnamefont {Wrachtrup}},
  \bibinfo {author} {\bibfnamefont {A.}~\bibnamefont {Pines}}, \ and\ \bibinfo
  {author} {\bibfnamefont {C.~A.}\ \bibnamefont {Meriles}},\ }\href {\doibase
  10.1103/PhysRevB.97.024422} {\bibfield  {journal} {\bibinfo  {journal} {Phys.
  Rev. B}\ }\textbf {\bibinfo {volume} {97}},\ \bibinfo {pages} {024422}
  (\bibinfo {year} {2018})}\BibitemShut {NoStop}%
\bibitem [{\citenamefont {Ajoy}\ \emph {et~al.}(2018)\citenamefont {Ajoy},
  \citenamefont {Nazaryan}, \citenamefont {Liu}, \citenamefont {Lv},
  \citenamefont {Safvati}, \citenamefont {Wang}, \citenamefont {Druga},
  \citenamefont {Reimer}, \citenamefont {Suter}, \citenamefont {Ramanathan},
  \citenamefont {Meriles},\ and\ \citenamefont {Pines}}]{AjoyDNP2018PNAS}%
  \BibitemOpen
  \bibfield  {author} {\bibinfo {author} {\bibfnamefont {A.}~\bibnamefont
  {Ajoy}}, \bibinfo {author} {\bibfnamefont {R.}~\bibnamefont {Nazaryan}},
  \bibinfo {author} {\bibfnamefont {K.}~\bibnamefont {Liu}}, \bibinfo {author}
  {\bibfnamefont {X.}~\bibnamefont {Lv}}, \bibinfo {author} {\bibfnamefont
  {B.}~\bibnamefont {Safvati}}, \bibinfo {author} {\bibfnamefont
  {G.}~\bibnamefont {Wang}}, \bibinfo {author} {\bibfnamefont {E.}~\bibnamefont
  {Druga}}, \bibinfo {author} {\bibfnamefont {J.~A.}\ \bibnamefont {Reimer}},
  \bibinfo {author} {\bibfnamefont {D.}~\bibnamefont {Suter}}, \bibinfo
  {author} {\bibfnamefont {C.}~\bibnamefont {Ramanathan}}, \bibinfo {author}
  {\bibfnamefont {C.~A.}\ \bibnamefont {Meriles}}, \ and\ \bibinfo {author}
  {\bibfnamefont {A.}~\bibnamefont {Pines}},\ }\href {\doibase
  10.1073/pnas.1807125115} {\bibfield  {journal} {\bibinfo  {journal}
  {Proceedings of the National Academy of Sciences}\ }\textbf {\bibinfo
  {volume} {115}},\ \bibinfo {pages} {10576} (\bibinfo {year}
  {2018})}\BibitemShut {NoStop}%
\bibitem [{\citenamefont {Jacques}\ \emph {et~al.}(2009)\citenamefont
  {Jacques}, \citenamefont {Neumann}, \citenamefont {Beck}, \citenamefont
  {Markham}, \citenamefont {Twitchen}, \citenamefont {Meijer}, \citenamefont
  {Kaiser}, \citenamefont {Balasubramanian}, \citenamefont {Jelezko},\ and\
  \citenamefont {Wrachtrup}}]{JacquesESLAC}%
  \BibitemOpen
  \bibfield  {author} {\bibinfo {author} {\bibfnamefont {V.}~\bibnamefont
  {Jacques}}, \bibinfo {author} {\bibfnamefont {P.}~\bibnamefont {Neumann}},
  \bibinfo {author} {\bibfnamefont {J.}~\bibnamefont {Beck}}, \bibinfo {author}
  {\bibfnamefont {M.}~\bibnamefont {Markham}}, \bibinfo {author} {\bibfnamefont
  {D.}~\bibnamefont {Twitchen}}, \bibinfo {author} {\bibfnamefont
  {J.}~\bibnamefont {Meijer}}, \bibinfo {author} {\bibfnamefont
  {F.}~\bibnamefont {Kaiser}}, \bibinfo {author} {\bibfnamefont
  {G.}~\bibnamefont {Balasubramanian}}, \bibinfo {author} {\bibfnamefont
  {F.}~\bibnamefont {Jelezko}}, \ and\ \bibinfo {author} {\bibfnamefont
  {J.}~\bibnamefont {Wrachtrup}},\ }\href {\doibase
  10.1103/PhysRevLett.102.057403} {\bibfield  {journal} {\bibinfo  {journal}
  {Phys. Rev. Lett.}\ }\textbf {\bibinfo {volume} {102}},\ \bibinfo {pages}
  {057403} (\bibinfo {year} {2009})}\BibitemShut {NoStop}%
\bibitem [{\citenamefont {Neumann}\ \emph {et~al.}(2010)\citenamefont
  {Neumann}, \citenamefont {Beck}, \citenamefont {Steiner}, \citenamefont
  {Rempp}, \citenamefont {Fedder}, \citenamefont {Hemmer}, \citenamefont
  {Wrachtrup},\ and\ \citenamefont {Jelezko}}]{JW2010Sci}%
  \BibitemOpen
  \bibfield  {author} {\bibinfo {author} {\bibfnamefont {P.}~\bibnamefont
  {Neumann}}, \bibinfo {author} {\bibfnamefont {J.}~\bibnamefont {Beck}},
  \bibinfo {author} {\bibfnamefont {M.}~\bibnamefont {Steiner}}, \bibinfo
  {author} {\bibfnamefont {F.}~\bibnamefont {Rempp}}, \bibinfo {author}
  {\bibfnamefont {H.}~\bibnamefont {Fedder}}, \bibinfo {author} {\bibfnamefont
  {P.~R.}\ \bibnamefont {Hemmer}}, \bibinfo {author} {\bibfnamefont
  {J.}~\bibnamefont {Wrachtrup}}, \ and\ \bibinfo {author} {\bibfnamefont
  {F.}~\bibnamefont {Jelezko}},\ }\href {\doibase 10.1126/science.1189075}
  {\bibfield  {journal} {\bibinfo  {journal} {Science}\ }\textbf {\bibinfo
  {volume} {329}},\ \bibinfo {pages} {542} (\bibinfo {year}
  {2010})}\BibitemShut {NoStop}%
\bibitem [{\citenamefont {Pagliero}\ \emph {et~al.}(2014)\citenamefont
  {Pagliero}, \citenamefont {Laraoui}, \citenamefont {Henshaw},\ and\
  \citenamefont {Meriles}}]{PaglieroAPL2014}%
  \BibitemOpen
  \bibfield  {author} {\bibinfo {author} {\bibfnamefont {D.}~\bibnamefont
  {Pagliero}}, \bibinfo {author} {\bibfnamefont {A.}~\bibnamefont {Laraoui}},
  \bibinfo {author} {\bibfnamefont {J.~D.}\ \bibnamefont {Henshaw}}, \ and\
  \bibinfo {author} {\bibfnamefont {C.~A.}\ \bibnamefont {Meriles}},\ }\href
  {\doibase 10.1063/1.4903799} {\bibfield  {journal} {\bibinfo  {journal}
  {Applied Physics Letters}\ }\textbf {\bibinfo {volume} {105}},\ \bibinfo
  {pages} {242402} (\bibinfo {year} {2014})}\BibitemShut {NoStop}%
\bibitem [{\citenamefont {Chakraborty}\ \emph {et~al.}(2017)\citenamefont
  {Chakraborty}, \citenamefont {Zhang},\ and\ \citenamefont
  {Suter}}]{Tanmoy2017pol}%
  \BibitemOpen
  \bibfield  {author} {\bibinfo {author} {\bibfnamefont {T.}~\bibnamefont
  {Chakraborty}}, \bibinfo {author} {\bibfnamefont {J.}~\bibnamefont {Zhang}},
  \ and\ \bibinfo {author} {\bibfnamefont {D.}~\bibnamefont {Suter}},\ }\href
  {http://stacks.iop.org/1367-2630/19/i=7/a=073030} {\bibfield  {journal}
  {\bibinfo  {journal} {New Journal of Physics}\ }\textbf {\bibinfo {volume}
  {19}},\ \bibinfo {pages} {073030} (\bibinfo {year} {2017})}\BibitemShut
  {NoStop}%
\bibitem [{\citenamefont {Xu}\ \emph {et~al.}(2019)\citenamefont {Xu},
  \citenamefont {Tian}, \citenamefont {Chen}, \citenamefont {Geng},
  \citenamefont {He}, \citenamefont {Wang},\ and\ \citenamefont
  {Du}}]{DuChopped2019}%
  \BibitemOpen
  \bibfield  {author} {\bibinfo {author} {\bibfnamefont {N.}~\bibnamefont
  {Xu}}, \bibinfo {author} {\bibfnamefont {Y.}~\bibnamefont {Tian}}, \bibinfo
  {author} {\bibfnamefont {B.}~\bibnamefont {Chen}}, \bibinfo {author}
  {\bibfnamefont {J.}~\bibnamefont {Geng}}, \bibinfo {author} {\bibfnamefont
  {X.}~\bibnamefont {He}}, \bibinfo {author} {\bibfnamefont {Y.}~\bibnamefont
  {Wang}}, \ and\ \bibinfo {author} {\bibfnamefont {J.}~\bibnamefont {Du}},\
  }\href {\doibase 10.1103/PhysRevApplied.12.024055} {\bibfield  {journal}
  {\bibinfo  {journal} {Phys. Rev. Applied}\ }\textbf {\bibinfo {volume}
  {12}},\ \bibinfo {pages} {024055} (\bibinfo {year} {2019})}\BibitemShut
  {NoStop}%
\bibitem [{\citenamefont {Dr\'eau}\ \emph {et~al.}(2012)\citenamefont
  {Dr\'eau}, \citenamefont {Maze}, \citenamefont {Lesik}, \citenamefont
  {Roch},\ and\ \citenamefont {Jacques}}]{Jacques13Chyperfine}%
  \BibitemOpen
  \bibfield  {author} {\bibinfo {author} {\bibfnamefont {A.}~\bibnamefont
  {Dr\'eau}}, \bibinfo {author} {\bibfnamefont {J.-R.}\ \bibnamefont {Maze}},
  \bibinfo {author} {\bibfnamefont {M.}~\bibnamefont {Lesik}}, \bibinfo
  {author} {\bibfnamefont {J.-F.}\ \bibnamefont {Roch}}, \ and\ \bibinfo
  {author} {\bibfnamefont {V.}~\bibnamefont {Jacques}},\ }\href {\doibase
  10.1103/PhysRevB.85.134107} {\bibfield  {journal} {\bibinfo  {journal} {Phys.
  Rev. B}\ }\textbf {\bibinfo {volume} {85}},\ \bibinfo {pages} {134107}
  (\bibinfo {year} {2012})}\BibitemShut {NoStop}%
\bibitem [{\citenamefont {Dr\'eau}\ \emph {et~al.}(2013)\citenamefont
  {Dr\'eau}, \citenamefont {Spinicelli}, \citenamefont {Maze}, \citenamefont
  {Roch},\ and\ \citenamefont {Jacques}}]{Jacques2NspinIni}%
  \BibitemOpen
  \bibfield  {author} {\bibinfo {author} {\bibfnamefont {A.}~\bibnamefont
  {Dr\'eau}}, \bibinfo {author} {\bibfnamefont {P.}~\bibnamefont {Spinicelli}},
  \bibinfo {author} {\bibfnamefont {J.~R.}\ \bibnamefont {Maze}}, \bibinfo
  {author} {\bibfnamefont {J.-F.}\ \bibnamefont {Roch}}, \ and\ \bibinfo
  {author} {\bibfnamefont {V.}~\bibnamefont {Jacques}},\ }\href {\doibase
  10.1103/PhysRevLett.110.060502} {\bibfield  {journal} {\bibinfo  {journal}
  {Phys. Rev. Lett.}\ }\textbf {\bibinfo {volume} {110}},\ \bibinfo {pages}
  {060502} (\bibinfo {year} {2013})}\BibitemShut {NoStop}%
\bibitem [{\citenamefont {Aslam}\ \emph {et~al.}(2013)\citenamefont {Aslam},
  \citenamefont {Waldherr}, \citenamefont {Neumann}, \citenamefont {Jelezko},\
  and\ \citenamefont {Wrachtrup}}]{Nabeel_Ion}%
  \BibitemOpen
  \bibfield  {author} {\bibinfo {author} {\bibfnamefont {N.}~\bibnamefont
  {Aslam}}, \bibinfo {author} {\bibfnamefont {G.}~\bibnamefont {Waldherr}},
  \bibinfo {author} {\bibfnamefont {P.}~\bibnamefont {Neumann}}, \bibinfo
  {author} {\bibfnamefont {F.}~\bibnamefont {Jelezko}}, \ and\ \bibinfo
  {author} {\bibfnamefont {J.}~\bibnamefont {Wrachtrup}},\ }\href
  {http://stacks.iop.org/1367-2630/15/i=1/a=013064} {\bibfield  {journal}
  {\bibinfo  {journal} {New Journal of Physics}\ }\textbf {\bibinfo {volume}
  {15}},\ \bibinfo {pages} {013064} (\bibinfo {year} {2013})}\BibitemShut
  {NoStop}%
\bibitem [{\citenamefont {Shin}\ \emph {et~al.}(2014)\citenamefont {Shin},
  \citenamefont {Butler}, \citenamefont {Wang}, \citenamefont {Avalos},
  \citenamefont {Seltzer}, \citenamefont {Liu}, \citenamefont {Pines},\ and\
  \citenamefont {Bajaj}}]{Bajaj14Nquadrapole}%
  \BibitemOpen
  \bibfield  {author} {\bibinfo {author} {\bibfnamefont {C.~S.}\ \bibnamefont
  {Shin}}, \bibinfo {author} {\bibfnamefont {M.~C.}\ \bibnamefont {Butler}},
  \bibinfo {author} {\bibfnamefont {H.-J.}\ \bibnamefont {Wang}}, \bibinfo
  {author} {\bibfnamefont {C.~E.}\ \bibnamefont {Avalos}}, \bibinfo {author}
  {\bibfnamefont {S.~J.}\ \bibnamefont {Seltzer}}, \bibinfo {author}
  {\bibfnamefont {R.-B.}\ \bibnamefont {Liu}}, \bibinfo {author} {\bibfnamefont
  {A.}~\bibnamefont {Pines}}, \ and\ \bibinfo {author} {\bibfnamefont {V.~S.}\
  \bibnamefont {Bajaj}},\ }\href {\doibase 10.1103/PhysRevB.89.205202}
  {\bibfield  {journal} {\bibinfo  {journal} {Phys. Rev. B}\ }\textbf {\bibinfo
  {volume} {89}},\ \bibinfo {pages} {205202} (\bibinfo {year}
  {2014})}\BibitemShut {NoStop}%
\bibitem [{\citenamefont {He}\ \emph {et~al.}(1993)\citenamefont {He},
  \citenamefont {Manson},\ and\ \citenamefont {Fisk}}]{Mansion14Nhyperfine}%
  \BibitemOpen
  \bibfield  {author} {\bibinfo {author} {\bibfnamefont {X.-F.}\ \bibnamefont
  {He}}, \bibinfo {author} {\bibfnamefont {N.~B.}\ \bibnamefont {Manson}}, \
  and\ \bibinfo {author} {\bibfnamefont {P.~T.~H.}\ \bibnamefont {Fisk}},\
  }\href {\doibase 10.1103/PhysRevB.47.8816} {\bibfield  {journal} {\bibinfo
  {journal} {Phys. Rev. B}\ }\textbf {\bibinfo {volume} {47}},\ \bibinfo
  {pages} {8816} (\bibinfo {year} {1993})}\BibitemShut {NoStop}%
\bibitem [{\citenamefont {Felton}\ \emph {et~al.}(2009)\citenamefont {Felton},
  \citenamefont {Edmonds}, \citenamefont {Newton}, \citenamefont {Martineau},
  \citenamefont {Fisher}, \citenamefont {Twitchen},\ and\ \citenamefont
  {Baker}}]{Felton2009}%
  \BibitemOpen
  \bibfield  {author} {\bibinfo {author} {\bibfnamefont {S.}~\bibnamefont
  {Felton}}, \bibinfo {author} {\bibfnamefont {A.~M.}\ \bibnamefont {Edmonds}},
  \bibinfo {author} {\bibfnamefont {M.~E.}\ \bibnamefont {Newton}}, \bibinfo
  {author} {\bibfnamefont {P.~M.}\ \bibnamefont {Martineau}}, \bibinfo {author}
  {\bibfnamefont {D.}~\bibnamefont {Fisher}}, \bibinfo {author} {\bibfnamefont
  {D.~J.}\ \bibnamefont {Twitchen}}, \ and\ \bibinfo {author} {\bibfnamefont
  {J.~M.}\ \bibnamefont {Baker}},\ }\href {\doibase 10.1103/PhysRevB.79.075203}
  {\bibfield  {journal} {\bibinfo  {journal} {Phys. Rev. B}\ }\textbf {\bibinfo
  {volume} {79}},\ \bibinfo {pages} {075203} (\bibinfo {year}
  {2009})}\BibitemShut {NoStop}%
\bibitem [{\citenamefont {Chen}\ \emph {et~al.}(2015)\citenamefont {Chen},
  \citenamefont {Hirose},\ and\ \citenamefont {Cappellaro}}]{PaolaPRB2015}%
  \BibitemOpen
  \bibfield  {author} {\bibinfo {author} {\bibfnamefont {M.}~\bibnamefont
  {Chen}}, \bibinfo {author} {\bibfnamefont {M.}~\bibnamefont {Hirose}}, \ and\
  \bibinfo {author} {\bibfnamefont {P.}~\bibnamefont {Cappellaro}},\ }\href
  {\doibase 10.1103/PhysRevB.92.020101} {\bibfield  {journal} {\bibinfo
  {journal} {Phys. Rev. B}\ }\textbf {\bibinfo {volume} {92}},\ \bibinfo
  {pages} {020101(R)} (\bibinfo {year} {2015})}\BibitemShut {NoStop}%
\bibitem [{\citenamefont {Shim}\ \emph
  {et~al.}(2013{\natexlab{b}})\citenamefont {Shim}, \citenamefont {Nowak},
  \citenamefont {Niemeyer}, \citenamefont {Zhang}, \citenamefont {Brandao},\
  and\ \citenamefont {Suter}}]{ShimArXiv}%
  \BibitemOpen
  \bibfield  {author} {\bibinfo {author} {\bibfnamefont {J.~H.}\ \bibnamefont
  {Shim}}, \bibinfo {author} {\bibfnamefont {B.}~\bibnamefont {Nowak}},
  \bibinfo {author} {\bibfnamefont {I.}~\bibnamefont {Niemeyer}}, \bibinfo
  {author} {\bibfnamefont {J.}~\bibnamefont {Zhang}}, \bibinfo {author}
  {\bibfnamefont {F.~D.}\ \bibnamefont {Brandao}}, \ and\ \bibinfo {author}
  {\bibfnamefont {D.}~\bibnamefont {Suter}},\ }\href
  {http://arxiv.org/abs/1307.0257} {\bibfield  {journal} {\bibinfo  {journal}
  {arXiv:1307.0257 [quant-ph]}\ } (\bibinfo {year}
  {2013}{\natexlab{b}})}\BibitemShut {NoStop}%
\bibitem [{\citenamefont {Rao}\ and\ \citenamefont
  {Suter}(2016)}]{Koti2016PRB}%
  \BibitemOpen
  \bibfield  {author} {\bibinfo {author} {\bibfnamefont {K.~R.~K.}\
  \bibnamefont {Rao}}\ and\ \bibinfo {author} {\bibfnamefont {D.}~\bibnamefont
  {Suter}},\ }\href {\doibase 10.1103/PhysRevB.94.060101} {\bibfield  {journal}
  {\bibinfo  {journal} {Phys. Rev. B}\ }\textbf {\bibinfo {volume} {94}},\
  \bibinfo {pages} {060101(R)} (\bibinfo {year} {2016})}\BibitemShut {NoStop}%
\bibitem [{\citenamefont {Simmons}\ \emph {et~al.}(2011)\citenamefont
  {Simmons}, \citenamefont {Brown}, \citenamefont {Riemann}, \citenamefont
  {Abrosimov}, \citenamefont {Becker}, \citenamefont {Pohl}, \citenamefont
  {Thewalt}, \citenamefont {Itoh},\ and\ \citenamefont
  {Morton}}]{MortonEnt2011}%
  \BibitemOpen
  \bibfield  {author} {\bibinfo {author} {\bibfnamefont {S.}~\bibnamefont
  {Simmons}}, \bibinfo {author} {\bibfnamefont {R.~M.}\ \bibnamefont {Brown}},
  \bibinfo {author} {\bibfnamefont {H.}~\bibnamefont {Riemann}}, \bibinfo
  {author} {\bibfnamefont {N.~V.}\ \bibnamefont {Abrosimov}}, \bibinfo {author}
  {\bibfnamefont {P.}~\bibnamefont {Becker}}, \bibinfo {author} {\bibfnamefont
  {H.-J.}\ \bibnamefont {Pohl}}, \bibinfo {author} {\bibfnamefont {M.~L.~W.}\
  \bibnamefont {Thewalt}}, \bibinfo {author} {\bibfnamefont {K.~M.}\
  \bibnamefont {Itoh}}, \ and\ \bibinfo {author} {\bibfnamefont {J.~J.~L.}\
  \bibnamefont {Morton}},\ }\href {\doibase 10.1038/nature09696} {\bibfield
  {journal} {\bibinfo  {journal} {Nature}\ }\textbf {\bibinfo {volume} {470}},\
  \bibinfo {pages} {69} (\bibinfo {year} {2011})}\BibitemShut {NoStop}%
\bibitem [{\citenamefont {Waldherr}\ \emph {et~al.}(2011)\citenamefont
  {Waldherr}, \citenamefont {Neumann}, \citenamefont {Huelga}, \citenamefont
  {Jelezko},\ and\ \citenamefont {Wrachtrup}}]{WaldherrLG2011}%
  \BibitemOpen
  \bibfield  {author} {\bibinfo {author} {\bibfnamefont {G.}~\bibnamefont
  {Waldherr}}, \bibinfo {author} {\bibfnamefont {P.}~\bibnamefont {Neumann}},
  \bibinfo {author} {\bibfnamefont {S.~F.}\ \bibnamefont {Huelga}}, \bibinfo
  {author} {\bibfnamefont {F.}~\bibnamefont {Jelezko}}, \ and\ \bibinfo
  {author} {\bibfnamefont {J.}~\bibnamefont {Wrachtrup}},\ }\href {\doibase
  10.1103/PhysRevLett.107.090401} {\bibfield  {journal} {\bibinfo  {journal}
  {Phys. Rev. Lett.}\ }\textbf {\bibinfo {volume} {107}},\ \bibinfo {pages}
  {090401} (\bibinfo {year} {2011})}\BibitemShut {NoStop}%
\end{thebibliography}%

\end{document}